\documentclass[preprint,12pt]{elsarticle}

\usepackage{amssymb}
\usepackage{amsmath}

\usepackage{listings}
\usepackage{multicol}
\usepackage{multirow}
\usepackage{makecell}
\usepackage{xcolor}
\usepackage{arydshln}
\usepackage{verbatim}

\definecolor{codegreen}{rgb}{0,0.6,0}
\definecolor{codegray}{rgb}{0.5,0.5,0.5}
\definecolor{codepurple}{rgb}{0.58,0,0.82}
\definecolor{backcolour}{rgb}{0.95,0.95,0.92}

\lstdefinestyle{mystyle}{
    backgroundcolor=\color{backcolour},   
    commentstyle=\color{codegreen},
    keywordstyle=\color{magenta},
    numberstyle=\tiny\color{codegray},
    stringstyle=\color{codepurple},
    basicstyle=\ttfamily\footnotesize,
    breakatwhitespace=false,         
    breaklines=true,                 
    captionpos=b,                    
    keepspaces=true,                 
    numbers=left,                    
    numbersep=5pt,                  
    showspaces=false,                
    showstringspaces=false,
    showtabs=false,                  
    tabsize=2
}
\usepackage[toc,page]{appendix}

\journal{arXiv}

\newcommand\restr[2]{{
  \left.\kern-\nulldelimiterspace 
  #1
  \littletaller 
  \right|_{#2} 
  }}

\newcommand{\littletaller}{\mathchoice{\vphantom{\big|}}{}{}{}}

\begin{document}

\begin{frontmatter}

\title{Metric Framework of Coherent Activity Patterns Identification in Spiking Neuronal Networks}

\author[1]{Daniil Radushev}	
 \affiliation[1]{organization={Center for Cognition and Decision Making, HSE University},
             city={Moscow},
            country={Russia}}
	
	\author[2]{Olesia Dogonasheva}
	\affiliation[2]{organization = {Université Paris Cité, Institut Pasteur, AP-HP, Inserm, Fondation Pour l'Audition, Institut de l’Audition, IHU reConnect, F-75012}, city =  {Paris}, country = {France} }
	\author[3]{Boris Gutkin}
	\affiliation[3]{organization={Group of Neural Theory and LNC2 INSERM U960, École Normale Supérieure PSL* University}, city={Paris}, country={France}}
	
	\author[1]{Denis Zakharov}
	\ead{dgzakharov@hse.ru}

\begin{abstract}

Partial synchronization plays a crucial role in the functioning of neuronal networks: selective, coordinated activation of neurons enables information processing that flexibly adapts to a changing computational context. Since the structure of coherent activity patterns reflects the network’s current state, developing automated tools to identify them is a key challenge in neurodynamics. Existing methods for analyzing neuronal dynamics tend to focus on global characteristics of the network, such as its aggregated synchrony level. While this approach can distinguish between the network’s main dynamical states, it cannot reveal the localization or properties of distinct coherent patterns.

In this work, we propose a new perspective on neural dynamics analysis that enables the study of network coherence at the single-neuron scale. We interpret the network as a metric space of neurons and represent its instantaneous state as an activity function on that space. We identify specific coherent activity clusters as regions where the activity function exhibits spatial continuity. Each cluster’s activity is further characterized using the analytical properties of the activity function within that region. This approach yields a concise yet detailed algorithmic profile of the network’s activity patterns.

\end{abstract}


\end{frontmatter}

\newpage
\section{Introduction}
\label{introduction}

In recent decades, the study of neuronal network dynamics in the brain has become a key area within the field of interdisciplinary nonlinear science. Neuronal networks give rise to complex collective activity patterns that underpin cognitive processes. Understanding the mechanisms that produce these patterns can help bridge the gap between micro-scale (single-cell), meso-scale (circuit), and macro-scale (functional) levels of brain function.

Neurons transmit information by propagating action potentials, whose generation depends on cellular properties (see, e.g., \cite{IZHIKEVICH_BOOK}) and synaptic inputs. In turn, spatially and temporally organized spike patterns serve as signatures of function and computation in neuronal networks \cite{SPATIOTEMPORAL_CESSAC, SPATIOTEMPORAL_ALDWORTH}.

In the dynamics of neuronal networks, evolving partially synchronous regimes play a crucial role \cite{PARTIAL_SYNCHRONY_IMPORTANCE}. Indeed, information processing relies on complex functional relationships between neurons: subpopulations are selectively activated and may enter synchronized regimes, while the rest of the network remains asynchronous. Hence, it is important not only to detect coherent activity patterns, but also to characterize their internal dynamics. Crucially, the analysis must go beyond identifying fully synchronous clusters to capture more complex coordinated activity patterns—so-called “coherent states”—which may involve spike-time delays.

Early research focused primarily on the emergence of global synchronization and the mechanisms that maintain it (see, e.g., \cite{SYNCHRONIZATION_BOOK_PIKOVSKY}), driven by applications such as clock synchronization \cite{CLOCK_SYNCHRONIZAION_Kapitaniak}, power grid stability \cite{POWER_GRID_SYNCHRONIZAION_Sajadi}, and laser dynamics \cite{LASER_SYNCHRONIZATION_Furst}. Subsequent research explored chimera states, in which synchronous and asynchronous activity coexist within networks of identical elements with symmetrical connections (see, e.g., the overview in \cite{CHIMERAS_REVIEW_PARASTESH}). However, developing methods to characterize detailed neuronal activity patterns within partially synchronized regimes remains an open challenge.

Traditionally, dynamic regimes in neuronal networks are assessed by computing a coherence parameter (e.g., the Kuramoto order parameter \cite{KURAMOTO_MODEL_REVISITED}, $\chi^2$ \cite{CHI2_GOLOMB}, Strength of Incoherence \cite{SI_DM_GOPAL}, or Cross-Correlation Coefficient \cite{CORRELATION_STRELKOVA}). These values typically range from 0 (full asynchrony) to 1 (full synchrony), with intermediate values corresponding to chimera states. While providing valuable data about the overall network synchronization level, those global measures often fail to distinguish between different dynamical regimes—such as global synchronization, multicluster synchronization, traveling (phase) wave and (multi)chimera states.

This limitation has been addressed through parameter ensembling. For example, the Adaptive Coherence Measure (ACM) approach \cite{ACM_DOGONASHEVA} evaluates two parameters: the coherence value $R^2$, obtained by maximizing $\chi^2$ over time delays for individual neurons' membrane potentials, and $L$, the number of unique delays found during this optimization. Together, these parameters allow reliable discrimination among key regimes in networks with ring topology and non-local connections, which are considered in most synchronization studies \cite{MULTISTABILITY_DOGONASHEVA}. However, the problem of localizing and characterizing the internal structure of coherent activity patterns remains unresolved.

In this work, we propose a new perspective on the analysis of neuronal network activity that enables localization and detailed description of coherent patterns. We frame the task as a clustering problem: which subsets of neurons exhibit coordinated activity? Our goal is to partition the network into regions of coherent activity and the remaining incoherent background. Once coherent regions are identified, we examine their internal structure by analyzing spatial variations in neuronal activity.

We introduce the Metric Framework (MF) for identifying coherent activity patterns in neuronal networks. This approach relies on a metric interpretation of the network. Each neuronal network possesses a spatial organization—either derived from its physical layout or inferred from the topology of synaptic connections. Based on this structure, a distance function can be defined between neurons, turning the neuron set into a metric space.

At a fixed time point $t_0$, we define the Activity Function (AF), which maps each neuron to a relevant activity characteristic (e.g., membrane potential) at that time. Regions where the AF changes continuously—i.e., where neighboring neurons have similar activity—are considered coherence clusters.\footnote{The idea of detecting coherence borders via discontinuities in neuronal activity was briefly mentioned in \cite{SI_DM_GOPAL}, but not developed into a systematic framework.} Within each cluster, the AF is a continuous function, and its analytic properties are used to classify the type of local coherence.

In this paper, we lay out the theoretical basis of the Metric Framework and provide guidelines for its application in neurodynamic studies. The Methods section introduces the key definitions and outlines the main analytical steps. The Results section summarizes the analysis pipeline and demonstrates its application on a symmetric ring network previously studied in \cite{RING_CALIM}. In the Discussion, we conclude the work and outline directions for future research.

\section{Methods} \label{sec2} 
\label{Methods}
\subsection{Main definitions}
\label{definitons}
\noindent The Metric Framework is based on three key notions: metric neuronal network, activity function, and coherent cluster. Next, we introduce their definitions. 
\\
\\
\noindent\textit{Definition 1:} \textbf{Metric neuronal network} (MNN) is a neuronal network, whose set of neurons $\mathcal{N}$ is equipped with a symmetric nonnegative \textit{distance function} $\rho: \mathcal{N}\times \mathcal{N} \rightarrow \mathbb{R}_{+}$ satisfying the triangle inequality and $\rho(\nu, \nu^\prime) = 0 \Leftrightarrow \nu = \nu ^\prime$.
\\
\\
For convenience, further in the article we refer to an MNN with the neuron set $\mathcal{N}$ simply as $\mathcal{N}$. The instantaneous state of an MNN at $t_0$ is encoded with an activity function mapping each neuron of the network to its current activity marker:
\\
\\
\textit{Definition 2:} Let $(X, d)$ be a metric space whose point set $X$ consists of possible neuron activity marker values\footnote{$X = \mathbb{R}$ if the membrane potential is used to characterize the instantaneous state of a single neuron; $X = S^1 = [0, 2\pi]_{/{0 \equiv 2\pi}}$ if the spiking phase is used as the neuron state marker,
etc.}, and $d$ is the distance function on $X$. Then, for a metric neuronal network $\mathcal{N}$ and a fixed timestamp $t_0$, a function $\mathrm{A}_{t_0}: \mathcal{N} \rightarrow X$ mapping each neuron $\nu \in \mathcal{N}$ to its instantaneous state marker at $t_0$ is called an \textbf{activity function}.
\\

Different activity marker choices give rise to different activity functions. For example, the \textbf{membrane potential function} mapping a neuron $\nu$ to its membrane potential $V_{t_0}(\nu)$ at the fixed $t_0$ is a $\mathbb{R}$-valued activity function. Another activity function choice is the \textbf{phase function} $\Phi_{t_0}$ mapping a neuron to its linear phase of spike generation $\phi_{\nu}(t_0)$:
\begin{equation}\Phi_{t_0}:\;\; \nu \mapsto \phi_\nu(t_0) = 2\pi \frac{t_0 - t_{\text{previous spike}}(\nu, t_0)}{t_{\text{next spike}}(\nu, t_0) - t_{\text{previous spike}}(\nu, t_0)},\end{equation}
which takes values in the phase circle $S^1 = [0, 2\pi]_{/{0 \equiv 2\pi}}$. In the following method description, we assume that the choice of activity function is fixed and refer to the activity function (AF). 

\begin{figure}[t]
\centering
\includegraphics[width = \linewidth]{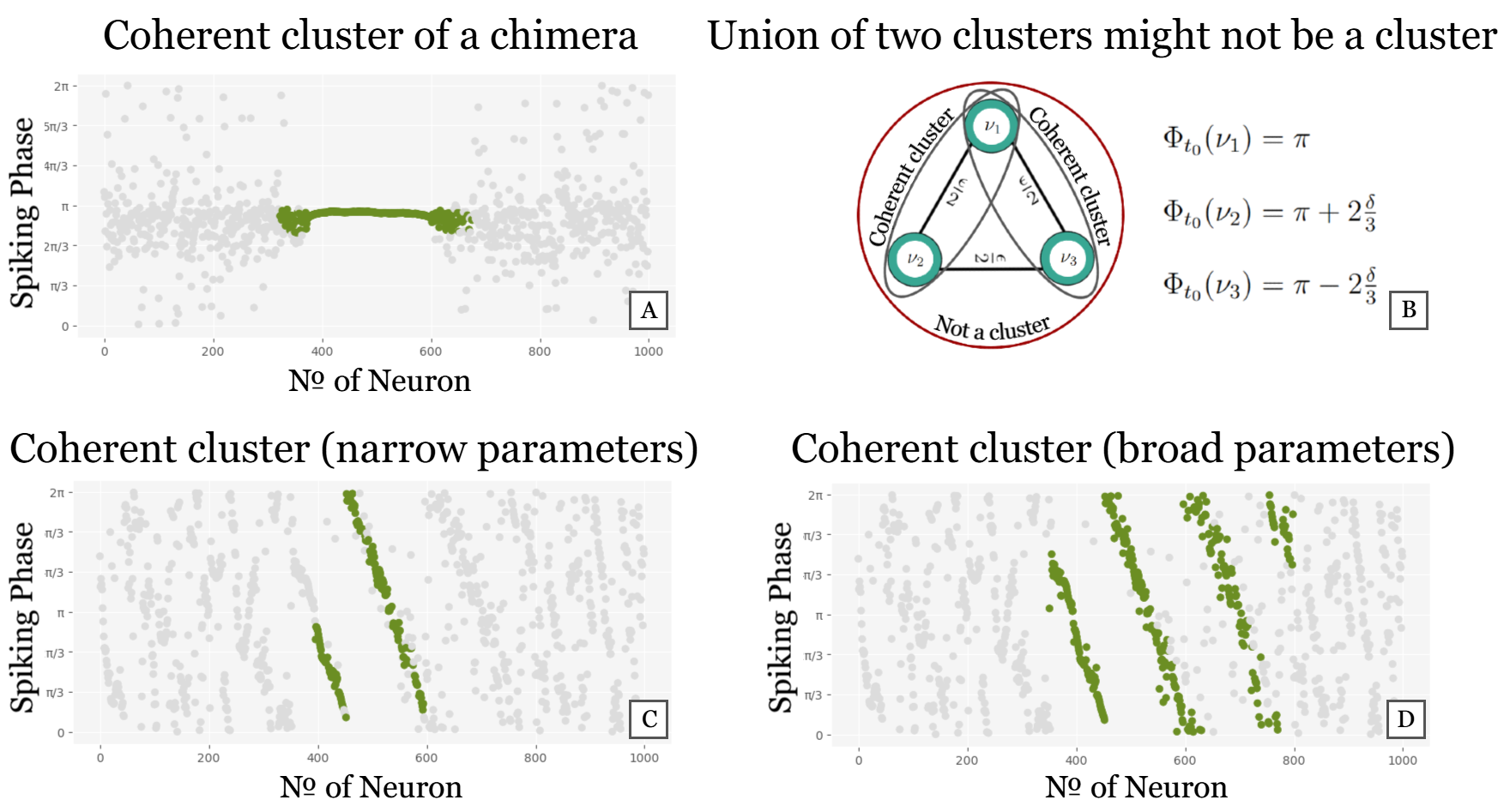}
\caption[Caption for LOF]{ Illustration of the notion of coherent cluster in the Metric Framework. (A) A coherent cluster (green points) is characterized by continuous phase function change, while in the incoherent region (gray points) the phase function is discontinuous. (B) An example of a network state in which a union of two coherent clusters is not a coherent cluster. Both $\{\nu_1,\nu_2 \}$ and $\{\nu_1, \nu_3 \}$ are coherent clusters, while $\{\nu_1, \nu_2, \nu_3\}$ is not. For that reason, there might be no \textit{largest} coherent cluster containing a given $\nu_1$. Nonetheless, there always exist clusters \textit{maximal} by inclusion: in this example, $\{\nu_1,\nu_2 \}$ and $\{\nu_1, \nu_3 \}$. (C,D) Coherent clusters identified according to the narrow and broad parameter choices are depicted. The narrow parameters configuration is more conservative, while the broad is more sensitive.}
\label{first_coherence_example}
\end{figure}
We define areas of the network with continuous change in the activity function between neighboring neurons as \textbf{coherent clusters} (see Figure \ref{first_coherence_example}A for an illustration). Since MNN is a discrete network, we use a finite \textbf{neuron proximity threshold} $\epsilon$ and an \textbf{activity proximity threshold} $\delta$ to evaluate the spatial continuity: a neuron set is a coherent cluster if $\epsilon$-close neurons have $\delta$-close activity parameter values. This concept is formalized in the following definition.
\\
\\
\textit{Definition 3:} A subset $\mathcal{C} \subset \mathcal{N}$ of neurons is called a ($\mathrm{A}_{t_0}, \epsilon, \delta$)-\textbf{coherent cluster} if the activity function $\mathrm{A}_{t_0}$ is $\delta$-close for $\epsilon$-neighbors within $\mathcal{C}$:
\[ \forall \nu, \nu^\prime \in \mathcal{C} \quad\quad \rho(\nu,\nu^\prime)\leq \epsilon \quad\Longrightarrow \quad d(\mathrm{A}_{t_0}(\nu), \mathrm{A}_{t_0}(\nu^\prime)) \;\leq \; \delta. \]

The proximity thresholds ($\epsilon,\delta$) can be tuned narrowly or broadly, depending on the desired balance between specificity and sensitivity. Under a narrow configuration ($\epsilon = 5$, $\delta = \frac{\pi}{10}$ in the ring-network example), the continuity criterion is stringent: only neurons that are almost contiguous and display nearly identical activity are assigned to the same coherent cluster (Fig.~\ref{first_coherence_example}C). This conservative setting minimizes false-positive detections of coherence, albeit at the cost of reduced sensitivity. Conversely, a broad configuration ($\epsilon = 15$, $\delta = \frac{3\pi}{10}$ in the ring network example) relaxes the continuity criterion. It allows neurons to be grouped together when they are separated by moderate distances and exhibiting moderately similar activity (Fig.~\ref{first_coherence_example}D). These permissive thresholds are advantageous for detecting spatially dispersed clusters that would be overlooked under the narrow configuration. Each parameter set (narrow and broad) highlights complementary aspects of network coherence. A comprehensive analysis should apply both configurations and compare the resulting cluster structures.

To automatize identification of the coherent clusters in the network, we employ graph algorithms. A coherent cluster is equivalently redefined in terms of $\epsilon$-neighbors graph. 
\\
\\
\noindent \textit{Definition 4:} For a subset $\mathcal{R} \subset \mathcal{N}$ of neurons of a MNN, the \textbf{Graph of $\epsilon-$neighbors} $G_\epsilon(\mathcal{R})$ is defined as the graph $(V, E)$ with $V = \mathcal{R}$ and 
\[ E = \{(\nu_1 \leftrightarrow \nu_2)\;\; |\;\; \nu_{1,2} \in \mathcal{R}, \;\; \rho(\nu_1, \nu_2) \leq \epsilon\}.\]

\noindent Then a ($\mathrm{A}_{t_0}, \epsilon, \delta$)-\textbf{coherent cluster} is a subset $\mathcal{C}\subset \mathcal{N}$ such that for any edge $(\nu_1 \leftrightarrow \nu_2)$ of the graph $G_\epsilon(\mathcal{C})$ the inequality $d(\mathrm{A}_{t_0}(\nu_1),\mathrm{A}_{t_0}(\nu_2)) \leq \delta$ holds. If the induced subgraph $G_\epsilon(\mathcal{C})$ of $G_{\epsilon}(\mathcal{N})$ on the vertex set $\mathcal{C}$ is connected, the cluster $\mathcal{C} \subset \mathcal{N}$ is called \textbf{$\epsilon$-connected}.
\subsection{Parsing network regions}
\label{region_parsing}
Within the metric framework, network-state analysis begins by locating the coherence clusters. The following theorem states that such a cluster containing a given neuron can be identified in linear time with respect to the size of the $\epsilon$-neighbors graph.
\\

\noindent\textbf{Theorem (the coherent cluster identification algorithm).} For a given neuron $\nu_1$, there exists an algorithm with running time $O(|V_{G_{\epsilon}(\mathcal{N})}| + |E_{G_{\epsilon}(\mathcal{N})}|)$ that returns an inclusion-maximal $\epsilon$-connected $(\mathrm{A}_{t_0}, \epsilon, \delta)$-coherent cluster containing $\nu_1$. 
\\

Notice that the theorem does not guarantee to find \textit{the largest} coherent cluster, since it might not exist (see Figure \ref{first_coherence_example}B for an illustration). The proof of the theorem (the algorithm pseudocode and the proof of its correctness) is provided in \ref{supplementary}.

\begin{figure}[t]
\centering
\includegraphics[width = \linewidth]{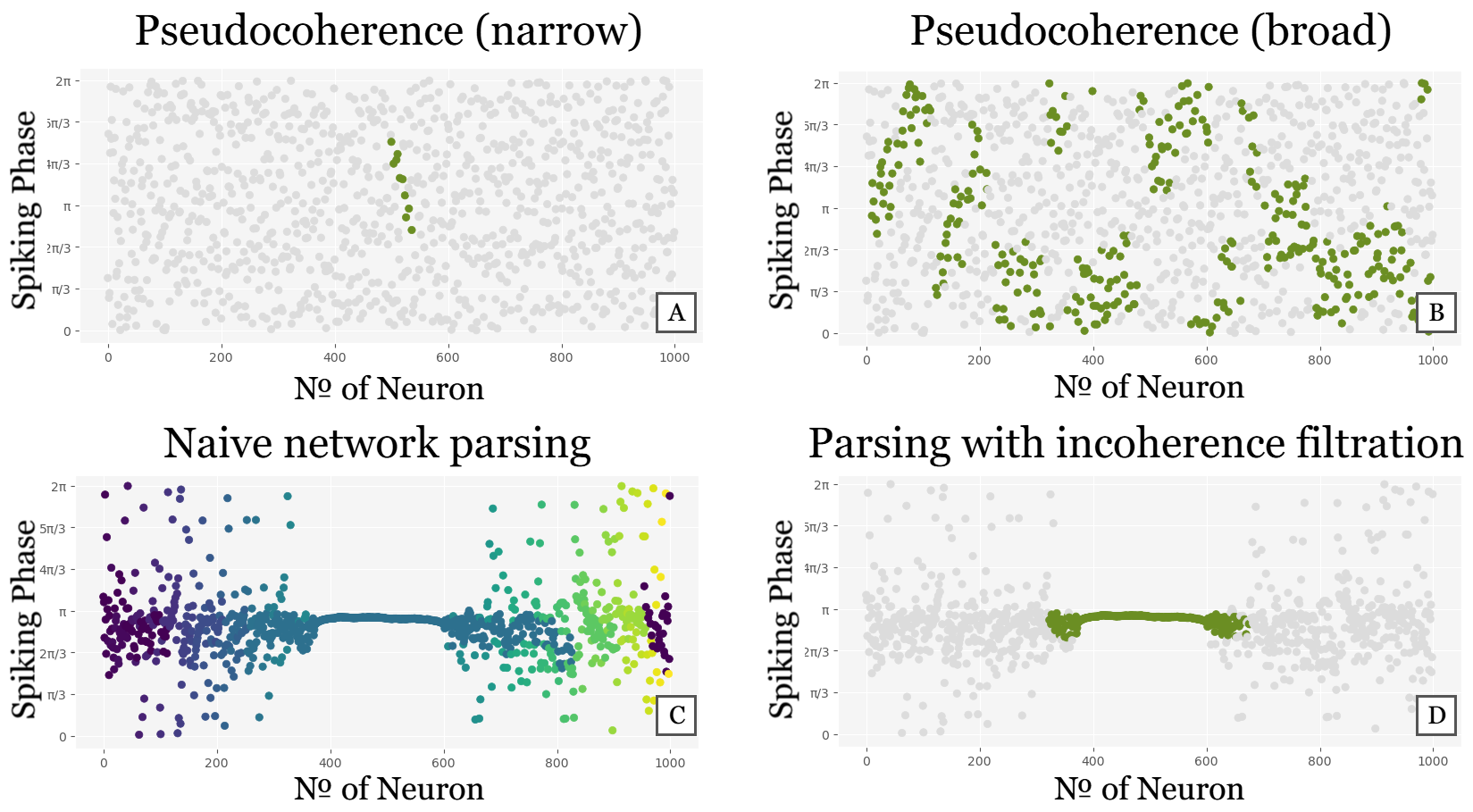}
\caption[Caption for LOF]{ Illustration of the pseudocoherence phenomenon and the challenges it poses. (A) A small pseudocoherent cluster of contingent neurons with incidentally similar activity that formally satisfies the \textit{Definiton 3} with narrow proximity thresholds. (B) A larger cluster identified according to broad parameters. (C) The result of network parsing without incoherence filtration is depicted: incoherence areas get divided into multiple pseudocoherent clusters. (D) Parsing with filtration. As a result of filtering, a natural parsing is yielded: the single coherence cluster is correctly identified (green points), and the incoherence areas are located (gray points).}
\label{ParsingFigure}
\end{figure}

Let us comment on how this method should be practically applied. For a given neuron, the algorithm identifies an $\epsilon$-connected coherent cluster that contains it and is maximal by inclusion. When the objective is to partition the entire network, an obvious—but inefficient—strategy is to consecutively apply the algorithm to all neurons of the network. This introduces a difficulty: many neurons reside in regions that are not genuinely coherent. Initializing the algorithm from such a neuron can yield a \emph{pseudocoherent cluster}—a set of contingent incoherent neurons having similar activity variables by chance, not due to actual coherence, and thus satisfying Definition 3 only formally (Fig.~\ref{ParsingFigure}A,B).  In incoherent regions, the activity function (Definition 2) is random. Given a sufficiently dense population of neurons, random fluctuations inevitably produce groups with superficially similar activity profiles. When the local neuron density is high, these chance alignments can generate pseudocoherent clusters that appear extensive in both space and size (Fig.~\ref{ParsingFigure}B). Proper network parsing therefore requires additional criteria to discriminate genuine coherent clusters from such artifacts.

We can address the pseudocoherence issue by the process of \textbf{incoherence filtration}, which can be applied at two points in the workflow. A pre-identification filter discards neurons that are incoherent before starting coherent cluster identification; a post-identification filter is applied after clusters have been extracted, excising those found to be pseudocoherent. We now integrate both strategies into a three-stage pipeline:

\begin{enumerate}
    \item Pre-identification filtration. Apply a heuristic that removes the majority of incoherent neurons (an example of such a heuristic is described below).
    \item Cluster extraction. Iteratively run the coherent cluster identification algorithm on the remaining neurons, thereby partitioning them into disjoint clusters.
    \item Post-identification filtration. Remove any cluster whose size is below the threshold $\xi|\mathcal{N}|$.
\end{enumerate}

The design of the pre-identification filter warrants explicit specification. We employ a concise yet effective heuristic — RCC filtration — that follows directly from the formal definition of a coherent cluster. 

In this work, a coherent cluster is defined as a subset of an MNN such that $\epsilon$-close neurons have $\delta$-close AF values (Definition 3). We can reformulate it as follows: for any given $\nu$ in a coherent cluster, $100\%$ of its $\epsilon-$neighbors \textit{within the cluster} have $\delta$-close AF values. Since neurons of a coherent cluster might have neighbors outside the cluster, the fraction of adjacent neurons with similar activity can decline when the calculation is performed over the entire neighborhood.

Despite this decrease, neurons in coherent clusters still have a greater fraction of activity-proximate neighbors than incoherent neurons. This disparity enables an effective filtration strategy: remove those neurons whose neighborhoods contain only a chance-level proportion of activity-similar neurons, thereby excluding candidates unlikely to participate in genuine coherence.

This motivates the \textbf{Relaxed Continuity Coefficient pre-identification filter}. For a fixed radius $\alpha > 0$ and activity tolerance $\beta > 0$, define the Relaxed Continuity Coefficient $\mathrm{K}^\beta_\alpha(\nu)$ of a neuron $\nu$ by 

\begin{equation}
    \mathrm{K}^\beta_\alpha(\nu) = \frac{|\mu: d(\mathrm{A}_{t_0}(\mu), \mathrm{A}_{t_0}(\nu)) \leq \beta|}{|\mu: \rho(\mu, \nu) \leq \alpha|}.    
\end{equation}

Thus $\mathrm{K}^\beta_\alpha(\nu)$ is the proportion of $\alpha$-neighbors whose activity functions differ from that of $\nu$ by less than $\beta$. For the substantiality threshold $\gamma > 0$, filter out neurons $\nu$ having $ \mathrm{K}^\beta_\alpha(\nu)<\gamma$ as incoherent.

Selecting $\alpha > \epsilon$, we alleviate the effects of random AF value coincidence by providing us with a larger sample of neighbors. Correspondingly, the condition $\beta < \delta$ compensates for the relaxed requirement $\gamma < 1$ and thereby tightens the activity-similarity criterion; since not every neighbor must match, the tolerance window can be narrowed without losing true positives.

In the numerical experiments (Section \ref{example}) we doubled $\alpha = 2\epsilon$ and halved $\beta = \frac{\delta}{2}$ as the parameters for the RCC filtration. For the narrow parameters $(\epsilon = \frac{\alpha}{2} = 5, \; \delta = 2 \beta = \frac{\pi}{10})$ we use the substantiality threshold $\gamma = 0.5$; on the post-identification filtering stage, we eliminate clusters occupying less than $\xi = 0.02$ share of the network. For the more sensitive broad parameters $(\epsilon = \frac{\alpha}{2} = 15, \; \delta = 2\beta = \frac{3\pi}{10})$ we use the decreased substantiality threshold $\gamma = 0.3$; on the post-identification filtering stage, we eliminate clusters occupying less than $\xi = 0.05$ share of the network.

\subsection{Identifying coherence modes within regions}
\label{within_cluster_analysis}

The second step of network-state analysis entails a qualitative characterization of each localized cluster by examining analytic properties of the activity function when restricted to that cluster.
For a cluster $\mathcal{C} \subset\mathcal{N}$, define its \textbf{maximal activity divergence} by

\begin{equation} \Delta_{\mathrm{A}_{t_0}}(\mathcal{C}) = \max_{\nu, \mu \in \mathcal{C}} d(\mathrm{A}_{t_0}(\nu), \mathrm{A}_{t_0}(\mu)).\end{equation}

Evaluating $\Delta_{\mathrm{A}_{t_0}}(\mathcal{C})$ discriminates between two canonical states: the synphase synchrony and the traveling wave. Indeed, during synphase synchrony, all neurons share the same activity value, so $\Delta_{\mathrm{A}_{t_0}}(\mathcal{C}) = 0$. In contrast, activity function is not constant in the traveling wave case: $\Delta_{\mathrm{A}_{t_0}}(\mathcal{C}) \neq 0$. Particularly, if $A = \Phi_{t_0}$ is the phase function, it changes linearly along the traveling wave cluster, yielding $\Delta_{\mathrm{A}_{t_0}}(\mathcal{C}) = \pi$. 

\begin{figure}[t]
\centering
\includegraphics[width = \linewidth]{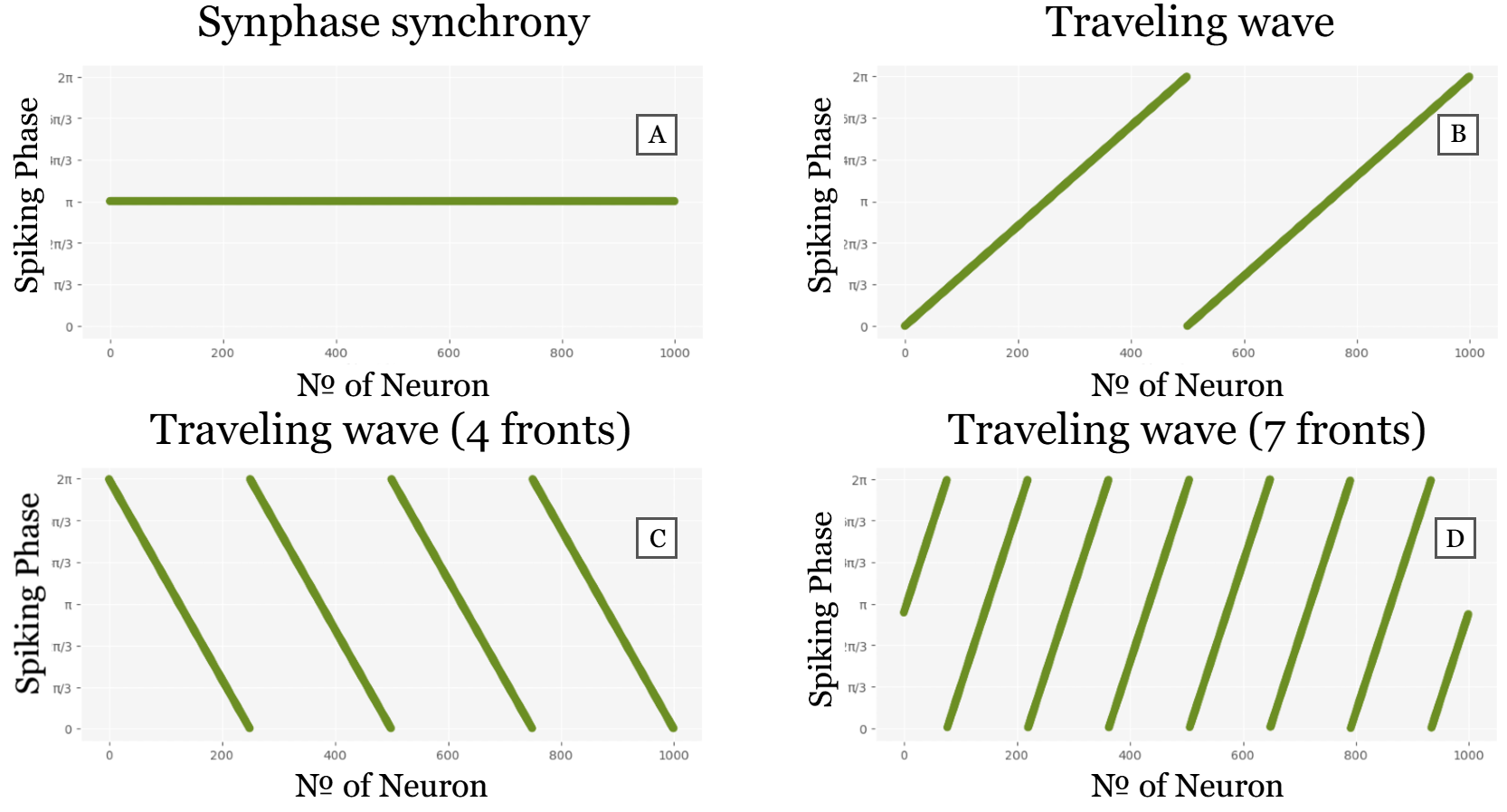}
\caption[Caption for LOF]{ Illustration of coherence mode identification using the activity function analysis. AF properties can be used not only to parse the network into coherence/incoherence clusters but also to characterize the coherence state within a single cluster. They allow both to distinguish between different major coherence types (A, B) and to finely characterize the state after its principal nature is deduced (traveling wave type difference (C, D)).} \label{withinclusterillustr}
\end{figure}

Once the coherence type of the cluster is identified, it can be described in further detail based on AF properties. For instance, the number of distinct wave fronts in a traveling wave corresponds to the number of full $2\pi$ phase windings of $\restr{\Phi_{t_0}}{\mathcal{C}}$. Formally, this is the cardinality of the preimage of $0$ on the phase circle $|\Phi_{t_0}^{-1}(0)|$ -- each element of which marks a set of neurons that spike simultaneously (see Fig.~\ref{withinclusterillustr}).

\section{Results}

In the Methods section, we introduced the Metric Framework of neural activity analysis. In the Results section, we proceed in two stages: first, we present a concise overview of the Metric Framework analysis pipeline; second, we apply this pipeline to distinguish the principal dynamic regimes in the ring network of I-type Morris–Lecar neurons studied in \cite{RING_CALIM}.

\subsection{Metric Framework: analysis pipeline}

The Metric Framework (MF) formalizes the local study of coherence within neuronal networks. Here, the network is considered as a metric space: each neuron is a point, and inter-neuron distances are specified by a chosen metric. At a fixed time $t_0$, we define the Activity Function (AF) that assigns to each neuron its instantaneous activity parameter (e.g., membrane potential or linear phase of spike generation). Regions in which AF changes continuously across adjacent neurons are designated as coherent clusters, while regions exhibiting irregular or unordered changes in AF are classified as incoherence areas and discarded. Once the clusters are located, their specific type of coherence -- such as synphase synchrony or traveling waves -- is determined by analyzing key analytic properties of AF (e.g., maximal activity divergence, phase-wrapping count).

\subsection{Example of implementation: main dynamic regimes of a ring network}
\label{example}

We illustrate the Metric Framework by classifying the dynamic regimes of the ring network studied in \cite{RING_CALIM}. This network comprises $N=1000$ I-type Morris-Lecar neurons \cite{MORRIS_LECAR}, each symmetrically connected to its $R$ nearest neighbors on either side via first-order kinetics excitatory synapses. The two principal control parameters are the synaptic strength $g$ and the connectivity density $r = \frac{R}{N}$ . For a detailed description of the model and its parameters, see the original paper \cite{RING_CALIM}. The distance between neurons is defined as the shortest path metric on the ring graph: $d(\nu_i, \nu_j) = \min(|i-j|, |\mathcal{N}| - |i-j|)$. The instantaneous state of the network is represented with the phase function—an activity function that uses the linear phase of spike generation as the single neuron state marker.

\renewcommand{\arraystretch}{3.9}
\begin{table}[t]
    \centering
    \caption{Classification of the main  states of the ring network from \cite{RING_CALIM}}
    \label{tab:table2}
    \begin{tabular}{|c|c|c|c|}
      \hline
      \multirowcell{2}{\textbf{Coherence type}\\ \textbf{exhibited by}\\ \textbf{coherent clusters}} & \multicolumn{3}{c|}{\textbf{Global coherence characteristic}}\\
      \cline{2-4}
      &  \makecell{\textbf{Full} \\ \textbf{Coherence}\\ \footnotesize{ No neurons} \\ \footnotesize{are filtered out} \\ \footnotesize{as incoherent}} & \makecell{\textbf{Chimera state}\\ \footnotesize{  Some neurons} \\ \footnotesize{are filtered out} \\ \footnotesize{as incoherent}}& \makecell{\textbf{Full} \\ \textbf{Incoherence} \\ \footnotesize{All neurons} \\ \footnotesize{are filtered out} \\ \footnotesize{as incoherent}}\\
      \hline
      \multirowcell{2}{\textbf{Synphase} \\ \textbf{synchrony} \\\\
      \footnotesize{Maximal Activity} \\
      \footnotesize{Divergence within the} \\
       \footnotesize{cluster is close to 0}}& \makecell{Global \\Synphase
       \\ Synchronization\\
       } & \makecell{Synphase
       \\ Chimera\\
       } & \multirowcell{4}{Incoherent\\State\\
       }\\ \cdashline{2-3}
     & \makecell{Multicluster\\ Synphase
     \\ Synchronization\\
     } & \makecell{Synphase
     \\ Multichimera\\
     } & \\ \cline{2-3}
      \cline{1-3}
    \multirowcell{2}{\textbf{Traveling Wave} \\\\
      \footnotesize{Maximal Activity} \\
      \footnotesize{Divergence within the} \\
       \footnotesize{cluster is significantly}
       \\ \footnotesize{greater than  0}}& \makecell{Traveling \\ Wave\\
       } & \makecell{Traveling Wave\\ Chimera\\
       } & \\ \cdashline{2-3}
       & \makecell{Traveling Waves \\ Superposition\\
       } & \makecell{Traveling Wave \\ Multichimera\\
       } & \\ \cline{2-3}
      \hline
    \end{tabular}
\end{table}

Previous studies distinguished regimes by the inspection of the aggregated coherence level of the network. For instance, in \cite{RING_CALIM}, the coherent state classification is based on the comparison of the SI parameter \cite{SI_DM_GOPAL} with a system of thresholds. By contrast, our approach performs a multi-aspect qualitative analysis of each instantaneous network state along three independent criteria: global coherence characteristic (fully coherent, chimera, or fully incoherent), number of coherent clusters, and type of intra-cluster coherence (synphase synchrony or traveling wave). Based on these three descriptors, we propose a classification of the states of the ring network in Table 1.

\begin{figure}[h]
\centering
\includegraphics[width = 1\linewidth]{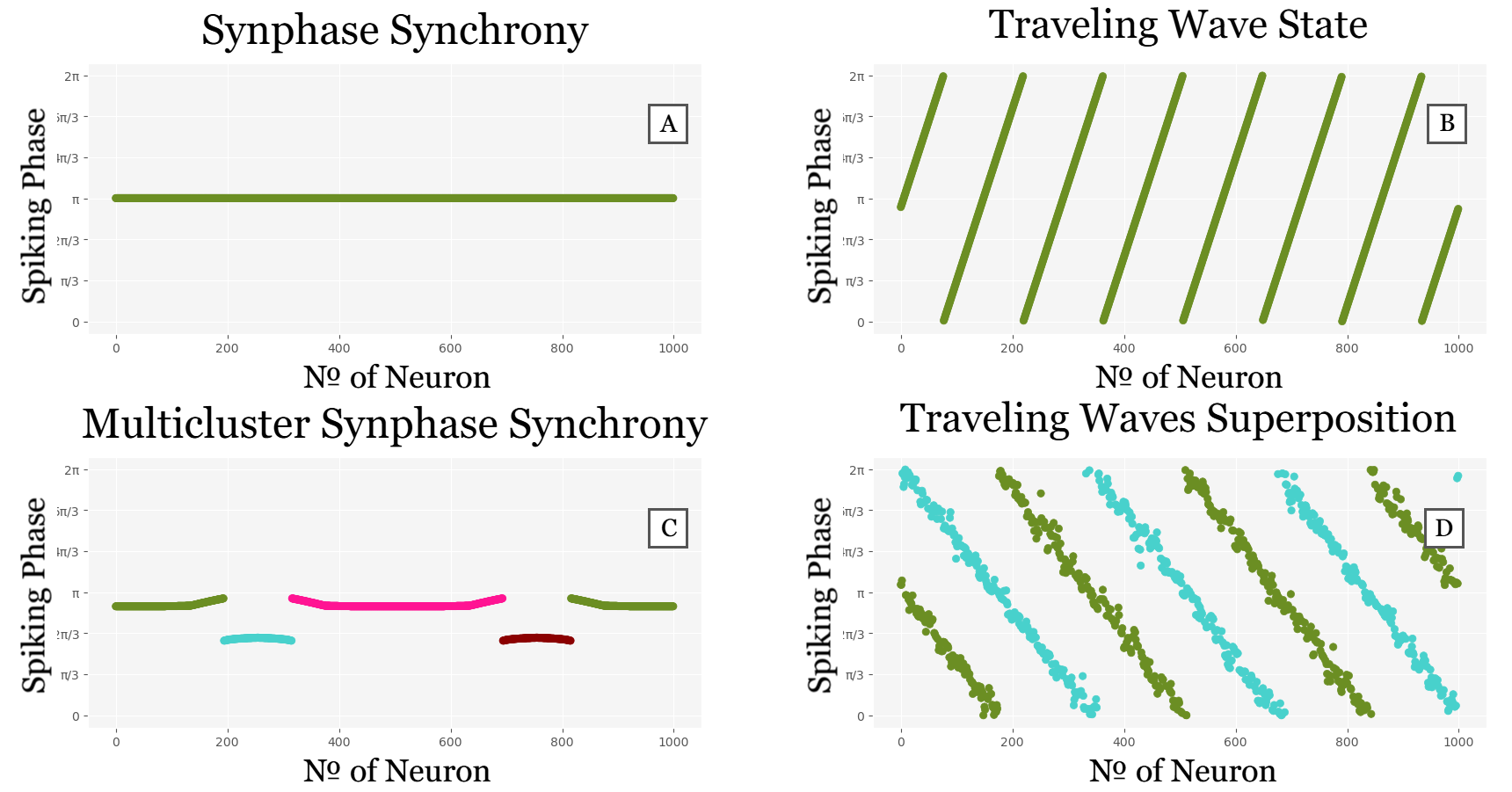}
\caption[Caption for LOF]{The four main coherent states of the ring network from \cite{RING_CALIM}. (A) Synphase synchronization (narrow parameters); (B) Traveling wave (narrow parameters); (C) Multiple clusters of synphase synchronization (narrow parameters); (D) Multiple clusters of traveling wave coherence (wide parameters).} \label{coherent_results}

\end{figure}

\begin{figure}[h]
\centering
\includegraphics[width = \linewidth]{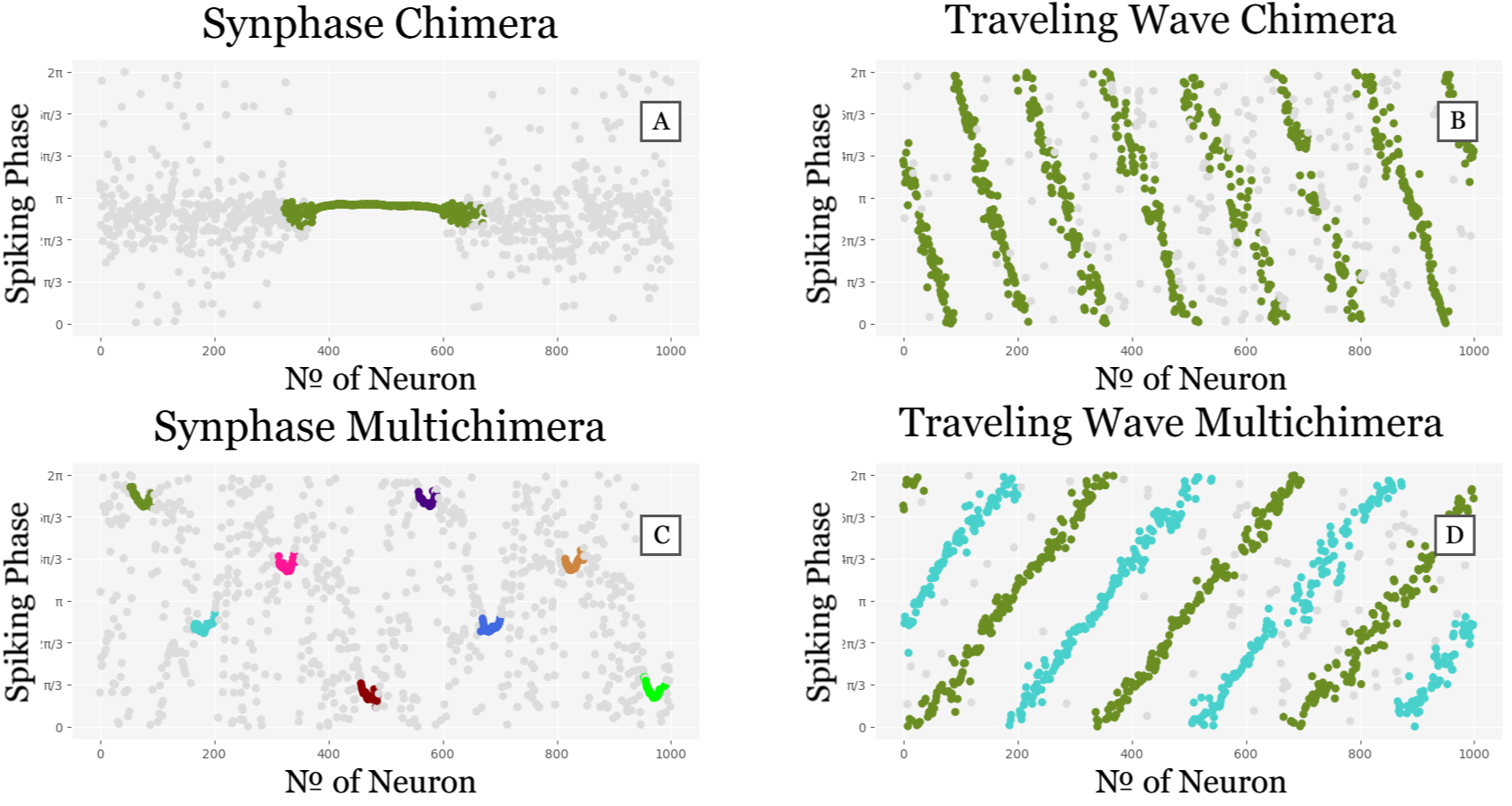}
\caption[Caption for LOF]{The four main chimera states of the ring network from \cite{RING_CALIM}. (A) Chimera with a synphase synchrony cluster (narrow parameters); (B) Chimera with a traveling wave cluster (wide parameters); (C) Multichimera with synphase synchronization clusters (narrow parameters); (D) Multichimera with traveling wave clusters (wide parameters). Gray points correspond to phases of incoherent neurons; brighter colors correspond to the phases of neurons in coherent clusters.
} \label{chimera_results}
\end{figure}

Fully coherent regimes arise when every neuron in the ring participates in some coherent cluster so that no cells are discarded as incoherent (Fig.~\ref{coherent_results}). In the simplest case, the phase function $\Phi_{t_0}$ is continuous around the entire ring, and the network forms one indivisible cluster. When $\Phi_{t_0}$ is constant (Fig.~\ref{coherent_results}A), the maximal phase divergence $\Delta_{\Phi_{t_0}} = 0$, and the network exhibit synphase synchrony. By contrast, if $\Phi_{t_0}$ increases monotonically along the ring—wrapping around the circle several times so that $\Delta_{\Phi_{t_0}} = \pi$ (Fig.~\ref{coherent_results}B)—the network has a traveling wave state. 

A more complex fully coherent scenario occurs when $\Phi_{t_0}$ is only piecewise continuous, breaking the ring into multiple disjoint segments of continuity. Figure~\ref{coherent_results}C shows four separate regions, each displaying negligible internal phase divergence and hence synchronous behavior within themselves, giving rise to a multicluster synphase synchrony. Figure~\ref{coherent_results}D shows two interlaced segments that each support a traveling wave ($\Delta_{\Phi_{t_0}} = \pi$); these waves are in antiphase and together constitute a superposition of traveling wave clusters.

The partially coherent chimera states are those in which coherent clusters border regions of discontinuous incoherent activity. Four types of  chimera states are illustrated in Figure~\ref{chimera_results}, where grey points mark filtered-out neurons. A synphase chimera consists of a single coherent cluster with nearly constant phase (so $\Delta_{\Phi_{t_0}} \approx 0$), embedded in an incoherent region (Fig.~\ref{chimera_results}A). If the phase function grows monotonously within the coherence cluster and $\Delta_{\Phi_{t_0}} = \pi$, the dynamical regime is a travelling wave chimera (Fig.~\ref{chimera_results}B). Beyond the single-cluster case, one can observe chimera patterns with multiple coherent clusters. When several synchronous clusters coexist amid incoherence (see Fig.~\ref{chimera_results}C, where $8$ continuous phase change regions within which $\Delta_{\Phi_{t_0}} \approx  0$), the network is in a synphase multichimera. When two antiphase travelling wave clusters are interspersed with incoherent neurons (Fig.~\ref{chimera_results}D), the network exhibits a travelling wave multichimera.

\begin{figure}[h]
\centering
\includegraphics[width = 0.75\linewidth]{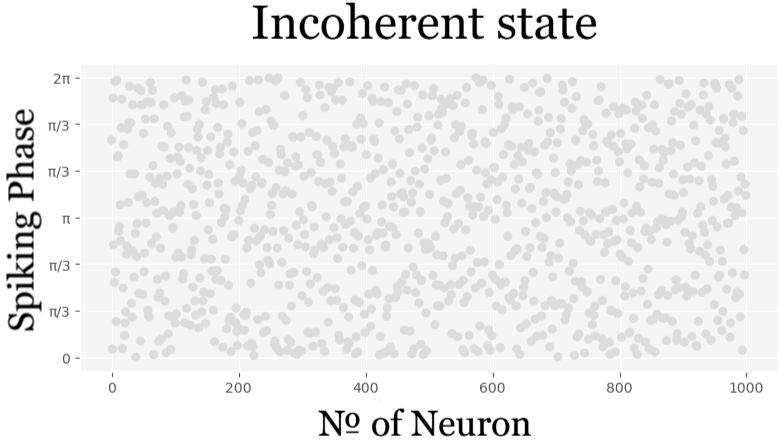}
\caption[Caption for LOF]{Incoherent state of the ring network from \cite{RING_CALIM}. This state is characterized by random-like phase function change: no coherent clusters are identified.} \label{incoherent_results}
\end{figure}

The incoherent regime is characterized by random-like change of the phase function: there are no regions of continuous phase change (Fig.~\ref{incoherent_results}). Thus, filtration removes every neuron as incoherent, and no coherent clusters can be identified.

\newpage
\section{Discussion}
\label{discussion}

The formation of coherent activity patterns in neuronal populations presents challenges to synchronization theory. These patterns are often involved in cognitive tasks and typically last for short durations \cite{SPATIOTEMPORAL_CESSAC, SPATIOTEMPORAL_ALDWORTH}. Furthermore, they do not involve all neurons in the network, but only the subsets required for specific computations.

Numerous approaches exist to classify synchronous regimes in neuronal networks using numerical parameters—or combinations of them—that characterize the network’s global state \cite{KURAMOTO_MODEL_REVISITED, CHI2_GOLOMB, SI_DM_GOPAL, CORRELATION_STRELKOVA, ACM_DOGONASHEVA, CORRELATION_DIMENSION_DOGONASHEVA, CHIMERA_SCHEMA}. However, these tools provide only a rough description of the network state and do not offer information about the localization or specific properties of distinct activity patterns. To facilitate more detailed investigation of neuronal network dynamics, there is a need for methods that focus on the characterization of individual patterns rather than the aggregated evaluation of the entire network.

In this paper, we introduce the Metric Framework (MF)—a novel approach to neuronal network activity analysis that enables the automatic localization and description of distinct coherent activity patterns at a given moment in time. This approach interprets the network as a metric space of neurons accompanied with an Activity Function (AF), which maps each neuron to its activity characteristic (e.g., membrane potential, spike phase) at a fixed time point $t_0$.

Coherent clusters are defined as regions where the AF changes continuously with respect to the network’s spatial structure, while abrupt changes in AF indicate incoherent regions. Within each coherent cluster, we analyze the analytic properties of the AF to determine the specific type of coherence it represents (e.g., synphase synchrony, traveling wave). In this way, the MF provides precise localization and characterization of coherent activity patterns.

Based on this information, one can also classify the network state: global coherence, where all neurons belong to some coherence cluster; incoherence, where no coherent clusters are present; and chimera states, in which coherent and incoherent regions coexist. Additionally, both globally coherent and chimera states can be further categorized by the number and types of coherent patterns present.

A core aspect of the MF is the spatial organization of neurons, which underpins the notion of spatial continuity. The choice of the metric structure—i.e., the distance function between neurons—is thus crucial. In networks with regular architecture (e.g., rings, grids), the natural metric is straightforward. For networks with moderately distorted regularity (e.g., Watts–Strogatz small-world networks \cite{SMALL_WORLD_WATTS_STROGATZ} with low-to-moderate $\beta$), one can use the metric from the undistorted topology (e.g., a ring metric). This is supported by results showing persistence of coherent activity patterns under gradual topological distortion \cite{CHIMERA_DEATH_ZAKHAROVA}.

More generally, each neuronal network possesses a spatial structure in the form of its connectivity graph, which can be used to infer the metric (e.g., the shortest path metric). Alternatively, one may define the metric inversely proportional to functional similarity between neurons. In future work, we plan to comparatively evaluate these options for networks with unclear or irregular spatial organization.

By shifting the focus from aggregated coherence measures to pattern localization, the MF connects analysis of neuronal network activity with fields such  as calculus on manifolds and classical machine learning. The link to calculus on manifolds arises because MF treats network state as a function over a metric space—analogous to scalar fields over manifolds—enabling analysis via concepts like continuity and smoothness. Although classical calculus cannot be directly applied due to the discrete nature of networks, we adapt its principles to work with the AF in this setting. 

The link to machine learning stems from the clustering task arising in our analysis: identification of spatially adjacent neuron sets with internally coordinated activity. While the problem of clustering in a metric space has been extensively studied in classical ML, standard tools cannot be applied in our context. Indeed, our request is conditional: identified clusters have to be characterized with continuous change of the activity function. Thus, an appropriate clustering procedure has to depend both on spatial and functional neuron proximity. In this work, we propose an algorithmic solution pipeline tailored to this domain-specific clustering problem.

A major future direction for the MF development is its generalization to dynamic analysis over time. In this article, we focused on a static snapshot of the network state represented with the current activity function. A natural extension is to study the temporal evolution of the AF, tracking the emergence, movement, transformation, and disappearance of coherent activity patterns. Such dynamic analysis could provide deeper insight into transient synchronization and its cognitive roles. Future work will focus on developing robust tools for analyzing time-parametrized activity functions.

\section{Acknowledgments}

This work/article is an output of a research project implemented as part of the Basic Research Program at the National Research University Higher School of Economics (HSE University). This research was supported in part through computational resources of HPC facilities at HSE University.
This work has benefited from a French government grant managed by the Agence Nationale de la Recherche under the France 2030 program, reference ANR-23-IAHU-0003. This work was supported by a grant from Fondation Pour l’Audition (FPA) to Anne-Lise Giraud (FPA IDA11).

\bibliographystyle{unsrt}
\bibliography{bib}

\newpage

\newpage
\appendix
\section{Proof of the Theorem}
\label{supplementary}
\subsection {Pseudocode implementing the cluster identification algorithm}
\begin{lstlisting}[language=Python, caption = Coherent cluster identification algorithm (python-style pseudocode)]
def GetMaxConnectedCluster(nu_1,   #Starting neuron
                             G,    #Epsilon-neighbors graph
                             A,    #Activity function
                             d):   #Metric on activity values
                             

    Q       = FIFO_Queue()  #Neurons to visit
    M       = set()         #Neurons added to the cluster
    C       = dict()        #Neurons' colors (visit statuses)
    Omega   = dict()        #Omega[nu] = Max(d(A(nu),A(mu))
                            # for mu in M and adjacent to nu)
    for nu in V(G):         
        C[nu] = White       #Initializing the colors and 
        Omega[nu] = 0       #the omega dictionary for all nu

    Q.enqueue(nu_1)         #Begining the traversal
    C[nu_1] = Gray          #starting from nu_1
    
    while Q is not empty:           
        nu = Q.dequeue()    #Take a neuron added to Q earlier
        C[nu] = Black       #Mark nu as visited
    
                               #Expand the cluster M if A(nu)
        if Omega[nu] <= delta: #is different from A of its
            M.add(nu)          #neighbors already added to M
                               #by no more than delta
            
            for mu adjacent to nu in G: #For nu's neighbors:
            
                if Omega[mu] < d(A(nu), A(mu)): #Update
                    Omega[mu] = d(A(nu), A(mu)) #Omega

                if C[mu] == White:              #Add mu to Q
                    Q.enqueue(mu)               #if not 
                    C[mu] = Gray                #enqueued yet

    return M


\end{lstlisting}
\subsection{Proof of the algorithm correctness}

 This algorithm is a version of breadth-first search traversal on $G_\epsilon(N)$, thus the complexity $O(|V_{G_\epsilon(N)}| + |E_{G_\epsilon(N)}|)$. In the algorithm, neurons are added to the cluster $M$ initiated with $\nu_1 \in M$. Neighbors of neurons already in $M$ are iteratively visited. If their activity value does not differ from neighbors in $M$ by more than $\delta$, they are added to the cluster $M$. The traversal is supported with the queue $Q$, the color dictionary $C$ and the dictionary $\Omega$. Color dictionary is used to track whether a neuron has been visited: neuron is colored black if it is already visited, gray if it is in the queue and white otherwise. Omega dictionary tracks the maximal difference between $A(\nu)$ and the activity of $\nu$'s neighbors already in $M$.

To proceed with the proof, let us introduce some notation. Denote as $\Theta(\nu)$ the set $\{ \mu : \rho(\mu, \nu ) \leq \epsilon \}$ of neurons adjacent to $\nu$ in $G_{\epsilon}(\mathcal{N})$. Let $Q_i$, $M_i$, $C_i$ and $\Omega_i$ be the states of the corresponding data structures after $i$ iterations of the "While" loop (lines 19-35), with $i = \infty$ denoting the final iteration. Let $\nu_i$ be the neuron dequeued (line 20) from $Q$ during the $i$th iteration of the while loop. If $\Omega_{i-1}[\nu] \leq \delta$, $\nu_i$ is added to $M_i$. To justify this check we prove the following lemma:
\\
\\
\textbf{Lemma.} For any neuron $\nu$ and index $i$ 
\[\Omega_i[\nu] = \max_{\nu^\prime \in M_i \cap \Theta(\nu)} d(\mathrm{A}_{t_0}(\nu), \mathrm{A}_{t_0}(\nu^\prime)),\]
where maximum over an empty set is defined as zero.
\\
\\
\textit{Proof of the lemma:} Induction on $i$. For $i=0$ the statement is trivial due to $M_0 = \emptyset$. Assume the statement is true for $i = k$. If $\nu_{k+1}$ does not pass the check on the line 24, neither $M$ nor $\Omega$ change: $\Omega_k = \Omega_{k+1}$, $M_k = M_{k+1}$, so the condition still holds for $i = k+1$. If $\nu_{k+1}$ passes the check, $M_{k+1} = M_{k} \sqcup \{\nu_{k+1}\}$. For $\mu \notin \Theta(\nu_{k+1})$, neither $\Omega[\mu]$ nor $M \cap \Theta(\mu)$ change. For $\mu $ adjacent to $\nu_{k+1}$ we have
\[ \Omega_{k+1}[\mu] = \max\big[\Omega_{k}[\mu], \;d(\mathrm{A}_{t_0}(\nu_{k+1}),\mathrm{A}_{t_0}(\mu))\big] = \]
\begin{equation}=  \max\big[\max_{\nu^\prime \in M_{k}\cap \Theta(\mu)} d(\mathrm{A}_{t_0}(\mu), \mathrm{A}_{t_0}(\nu^\prime)), \;\;d(\mathrm{A}_{t_0}(\mu), \mathrm{A}_{t_0}(\nu_{k+1}))\big]  = \end{equation} 
\[= \max_{\nu^\prime \in M_{k+1}\cap \Theta(\mu) } d(\mathrm{A}_{t_0}(\mu), \mathrm{A}_{t_0}(\nu^\prime)). \text{ \qed}\]
The lemma immediately infers: $M_i$ is a coherent cluster for all $i\in \mathbb{N}\cup\{\infty\}$. Indeed, $M_1 = \{\nu_1\}$ is a coherent cluster like any single-neuron set. Next, if $M_k$ is a coherent cluster and $\nu_{k+1}$ is added to $M_{k+1}$, then the $\epsilon,\delta-$coherence condition is not broken by $\mathrm{A}_{t_0}(\nu_{k+1})$:
\begin{equation} \max_{\nu^\prime \in M_k \cap \Theta(\nu_{k+1})} d(\mathrm{A}_{t_0}(\nu_{k+1}), \mathrm{A}_{t_0}(\nu^\prime))\leq \delta.\end{equation}
The cluster is $\epsilon$-connected: any neuron $\nu \in M_i$ has a path to $\nu_1$ in $G_{\epsilon}(M_i)$ by the traversal design.

Finally, let us prove by contradiction that $M = M_\infty$ is maximal. Suppose that there exists a connected coherent cluster $M^* \supsetneq M$. Then there exists a pair of adjacent neurons $\nu, \mu$ such that $\nu \in M, \;\mu \in M^* \setminus M$. Since $\nu \in M$, $\nu = \nu_i$ for $i$ such that $\nu \in M_{i} \setminus M_{i-1}$. The neuron $\mu$ is adjacent to $\nu_i$, so it was enqueued during the $i$th loop iteration or earlier and dequeued at some step $j$: $\mu = \nu_j$. Since on the $j$th  iteration $\mu = \nu_j$ was not added to $M_j$, $\Omega_{j-1}[\mu] > \delta$. From the lemma
\begin{equation} \exists \kappa \in M_{j-1} \cap \Theta(\mu): \;\; d(\mathrm{A}_{t_0}(\mu),\mathrm{A}_{t_0}(\kappa)) > \delta.\end{equation}
Since $M_{j-1} \subset M \subset M^*$, $\kappa \in M^*$. Thus, we have a pair of adjacent neurons $\mu, \kappa \in M^*$ having $d(\mathrm{A}_{t_0}(\mu),\mathrm{A}_{t_0}(\kappa)) > \delta$, which contradicts the assumption that $M^*$ is a coherent cluster. Therefore, there exists no connected coherent cluster properly containing $M$. The theorem is proven. \qed

\newpage

\end{document}